\documentstyle[twoside]{article}

\catcode`\@=11
\long\def\@makefntext#1{
\protect\noindent \hbox to 3.2pt {\hskip-.9pt  
$^{{\eightrm\@thefnmark}}$\hfil}#1\hfill}		

\def\@makefnmark{\hbox to 0pt{$^{\@thefnmark}$\hss}}	
	
\def\ps@myheadings{\let\@mkboth\@gobbletwo
\def\@oddhead{\hbox{}
\rightmark\hfil\eightrm\thepage}   
\def\@oddfoot{}\def\@evenhead{\eightrm\thepage\hfil
\leftmark\hbox{}}\def\@evenfoot{}
\def\sectionmark##1{}\def\subsectionmark##1{}}

\oddsidemargin=\evensidemargin
\newcounter{sectionc}\newcounter{subsectionc}\newcounter{subsubsectionc}
\renewcommand{\section}[1] {\vspace{12pt}\addtocounter{sectionc}{1} 
\setcounter{subsectionc}{0}\setcounter{subsubsectionc}{0}\noindent 
	{\tenbf\thesectionc. #1}\par\vspace{5pt}}
\renewcommand{\subsection}[1] {\vspace{12pt}\addtocounter{subsectionc}{1} 
	\setcounter{subsubsectionc}{0}\noindent 
	{\bf\thesectionc.\thesubsectionc. {\kern1pt \bfit #1}}\par\vspace{5pt}}
\renewcommand{\subsubsection}[1] {\vspace{12pt}\addtocounter{subsubsectionc}{1}
	\noindent{\tenrm\thesectionc.\thesubsectionc.\thesubsubsectionc.
	{\kern1pt \tenit #1}}\par\vspace{5pt}}
\newcommand{\nonumsection}[1] {\vspace{12pt}\noindent{\tenbf #1}
	\par\vspace{5pt}}

\newcounter{appendixc}
\newcounter{subappendixc}[appendixc]
\newcounter{subsubappendixc}[subappendixc]
\renewcommand{\thesubappendixc}{\Alph{appendixc}.\arabic{subappendixc}}
\renewcommand{\thesubsubappendixc}
	{\Alph{appendixc}.\arabic{subappendixc}.\arabic{subsubappendixc}}

\renewcommand{\appendix}[1] {\vspace{12pt}
        \refstepcounter{appendixc}
        \setcounter{figure}{0}
        \setcounter{table}{0}
        \setcounter{lemma}{0}
        \setcounter{theorem}{0}
        \setcounter{corollary}{0}
        \setcounter{definition}{0}
        \setcounter{equation}{0}
        \renewcommand{\thefigure}{\Alph{appendixc}.\arabic{figure}}
        \renewcommand{\thetable}{\Alph{appendixc}.\arabic{table}}
        \renewcommand{\theappendixc}{\Alph{appendixc}}
        \renewcommand{\thelemma}{\Alph{appendixc}.\arabic{lemma}}
        \renewcommand{\thetheorem}{\Alph{appendixc}.\arabic{theorem}}
        \renewcommand{\thedefinition}{\Alph{appendixc}.\arabic{definition}}
        \renewcommand{\thecorollary}{\Alph{appendixc}.\arabic{corollary}}
        \renewcommand{\theequation}{\Alph{appendixc}.\arabic{equation}}
        \noindent{\tenbf Appendix \theappendixc #1}\par\vspace{5pt}}
\newcommand{\subappendix}[1] {\vspace{12pt}
        \refstepcounter{subappendixc}
        \noindent{\bf Appendix \thesubappendixc. {\kern1pt \bfit #1}}
	\par\vspace{5pt}}
\newcommand{\subsubappendix}[1] {\vspace{12pt}
        \refstepcounter{subsubappendixc}
        \noindent{\rm Appendix \thesubsubappendixc. {\kern1pt \tenit #1}}
	\par\vspace{5pt}}

\newcommand{\textlineskip}{\baselineskip=13pt}
\newcommand{\smalllineskip}{\baselineskip=10pt}

\def\eightcirc{
\begin{picture}(0,0)
\put(4.4,1.8){\circle{6.5}}
\end{picture}}
\def\eightcopyright{\eightcirc\kern2.7pt\hbox{\eightrm c}} 

\newcommand{\copyrightheading}[1]
	{\vspace*{-2.5cm}\smalllineskip{\flushleft
	{\footnotesize International Journal of Modern Physics A, #1}\\
	{\footnotesize $\eightcopyright$\, World Scientific Publishing
	 Company}\\
	 }}

\def\abstracts#1#2#3{{
	\centering{\begin{minipage}{4.5in}\baselineskip=10pt\footnotesize
	\parindent=0pt #1\par 
	\parindent=15pt #2\par
	\parindent=15pt #3
	\end{minipage}}\par}} 

\renewenvironment{thebibliography}[1]
	{\frenchspacing
	 \ninerm\baselineskip=11pt
	 \begin{list}{\arabic{enumi}.}
	{\usecounter{enumi}\setlength{\parsep}{0pt}
	 \setlength{\leftmargin 12.7pt}{\rightmargin 0pt} 
	 \setlength{\itemsep}{0pt} \settowidth
	{\labelwidth}{#1.}\sloppy}}{\end{list}}

\newcounter{itemlistc}
\newcounter{romanlistc}
\newcounter{alphlistc}
\newcounter{arabiclistc}

\newcommand{\fcaption}[1]{
        \refstepcounter{figure}
        \setbox\@tempboxa = \hbox{\footnotesize Fig.~\thefigure. #1}
        \ifdim \wd\@tempboxa > 5in
           {\begin{center}
        \parbox{5in}{\footnotesize\smalllineskip Fig.~\thefigure. #1}
            \end{center}}
        \else
             {\begin{center}
             {\footnotesize Fig.~\thefigure. #1}
              \end{center}}
        \fi}

\newcommand{\tcaption}[1]{
        \refstepcounter{table}
        \setbox\@tempboxa = \hbox{\footnotesize Table~\thetable. #1}
        \ifdim \wd\@tempboxa > 5in
           {\begin{center}
        \parbox{5in}{\footnotesize\smalllineskip Table~\thetable. #1}
            \end{center}}
        \else
             {\begin{center}
             {\footnotesize Table~\thetable. #1}
              \end{center}}
        \fi}

\def\@citex[#1]#2{\if@filesw\immediate\write\@auxout
	{\string\citation{#2}}\fi
\def\@citea{}\@cite{\@for\@citeb:=#2\do
	{\@citea\def\@citea{,}\@ifundefined
	{b@\@citeb}{{\bf ?}\@warning
	{Citation `\@citeb' on page \thepage \space undefined}}
	{\csname b@\@citeb\endcsname}}}{#1}}

\newif\if@cghi
\def\cite{\@cghitrue\@ifnextchar [{\@tempswatrue
	\@citex}{\@tempswafalse\@citex[]}}
\def\citelow{\@cghifalse\@ifnextchar [{\@tempswatrue
	\@citex}{\@tempswafalse\@citex[]}}
\def\@cite#1#2{{$\null^{#1}$\if@tempswa\typeout
	{IJCGA warning: optional citation argument 
	ignored: `#2'} \fi}}

\def\pmb#1{\setbox0=\hbox{#1}
	\kern-.025em\copy0\kern-\wd0
	\kern.05em\copy0\kern-\wd0
	\kern-.025em\raise.0433em\box0}

\def\fnt#1#2{\footnotetext{\kern-.3em
	{$^{\mbox{\scriptsize #1}}$}{#2}}}

\def\fpage#1{\begingroup
\voffset=.3in
\thispagestyle{empty}\begin{table}[b]\centerline{\footnotesize #1}
	\end{table}\endgroup}

\def\runninghead#1#2{\pagestyle{myheadings}
\markboth{{\protect\footnotesize\it{\quad #1}}\hfill}
{\hfill{\protect\footnotesize\it{#2\quad}}}}
\font\tenrm=cmr10
\font\tenit=cmti10 
\font\tenbf=cmbx10
\font\bfit=cmbxti10 at 10pt
\font\ninerm=cmr9

\font\eightrm=cmr8

\def\qed{\hbox{${\vcenter{\vbox{			
   \hrule height 0.4pt\hbox{\vrule width 0.4pt height 6pt
   \kern5pt\vrule width 0.4pt}\hrule height 0.4pt}}}$}}

\def\d0{D_2^{*0}}
\def\d+{D_2^{*+}}
\def\mev{ \,{\rm MeV}/c^2  }
\begin{document}

\runninghead{S.~Sarwar (FOCUS Coll.), PRELIMINARY $\ldots$ CHARMED MESON SPECTROSCOPY
}  
            {S.~Sarwar (FOCUS Coll.), PRELIMINARY $\ldots$ CHARMED MESON SPECTROSCOPY
} 
\normalsize\textlineskip
\thispagestyle{empty}
\setcounter{page}{1}

\copyrightheading{}			

\vspace*{0.88truein}

\fpage{1}
\centerline{\bf PRELIMINARY  RESULTS ON CHARMED MESON SPECTROSCOPY }
\vspace*{0.37truein}
\centerline{\footnotesize SHAHZAD SARWAR }
\vspace*{0.015truein}
\centerline{\footnotesize\it Laboratori Nazionali di Frascati, v. E.Fermi,
40 - Frascati   - 
 00044 Italy  }
\vspace*{10pt}
\centerline{
  \footnotesize{
    on behalf of the FOCUS Collaboration
     \footnote{
        Coauthors: J.M. Link, V.S. Paolone, M. Reyes, P.M. Yager
        ({\bf UC DAVIS}); J.C. Anjos,
        I. Bediaga, C. G\"obel, J. Magnin, J.M. de Miranda, I.M. Pepe
% ...
}
    }
   }
\vspace*{0.225truein}
\abstracts{
We report the preliminary measurement by the FOCUS Collaboration
 (E831 at Fermilab) of masses and widths of the
L=1 charm mesons $D_2^{*0}$ and $D_2^{*+}$.
The fit of the invariant mass distribution
requires an additional  term to account for a broad structure over background.
}{}{}

\textlineskip			
\vspace*{12pt}			

\noindent
 We present preliminary results from the FOCUS experiment
 (E831 at Fermilab)
 on the spectroscopy of bound states of a charm quark and a light quark
 with orbital angular momentum  $L=1$, called $D_2^* (c\bar u, c\bar d)$.
 A  theoretical framework for the spectrum of
 heavy-light mesons is given by Heavy Quark Symmetry (HQS),
 which predicts both narrow and broad states.
 While the narrow states are
 well established, the evidence for the broad states (both in the $c$-quark
 and in the $b$-quark sector) is much less stringent \cite{reviews}.
%
%
\par
The data for this paper were collected in the Wideband photoproduction
experiment FOCUS during the Fermilab 1996--1997 fixed-target
run. FOCUS is a considerably upgraded version of a previous experiment,
E687 \cite{Frabetti:1992au}. In FOCUS, a forward multi-particle
spectrometer is used to
measure the interactions of high-energy photons on a segmented BeO
target. We obtained a sample of over 1 million fully reconstructed
charm particles in the three major decay modes: $D\rightarrow K\pi, K2\pi,
K3\pi$. 
%
%
\par
\begin{figure}[ht]
 \vspace{11.0cm}
  \includegraphics{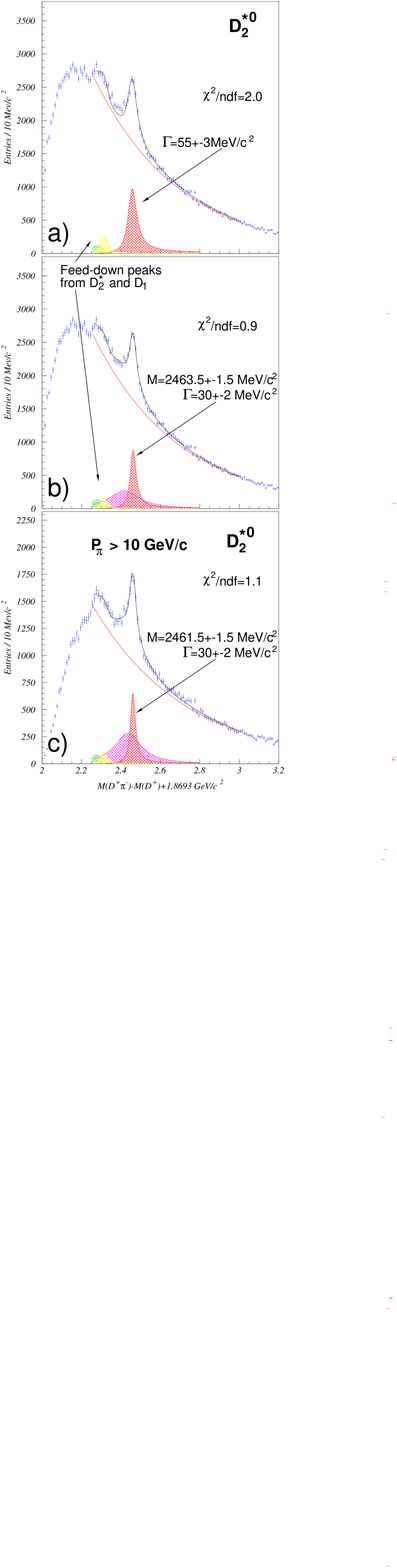}
  \includegraphics{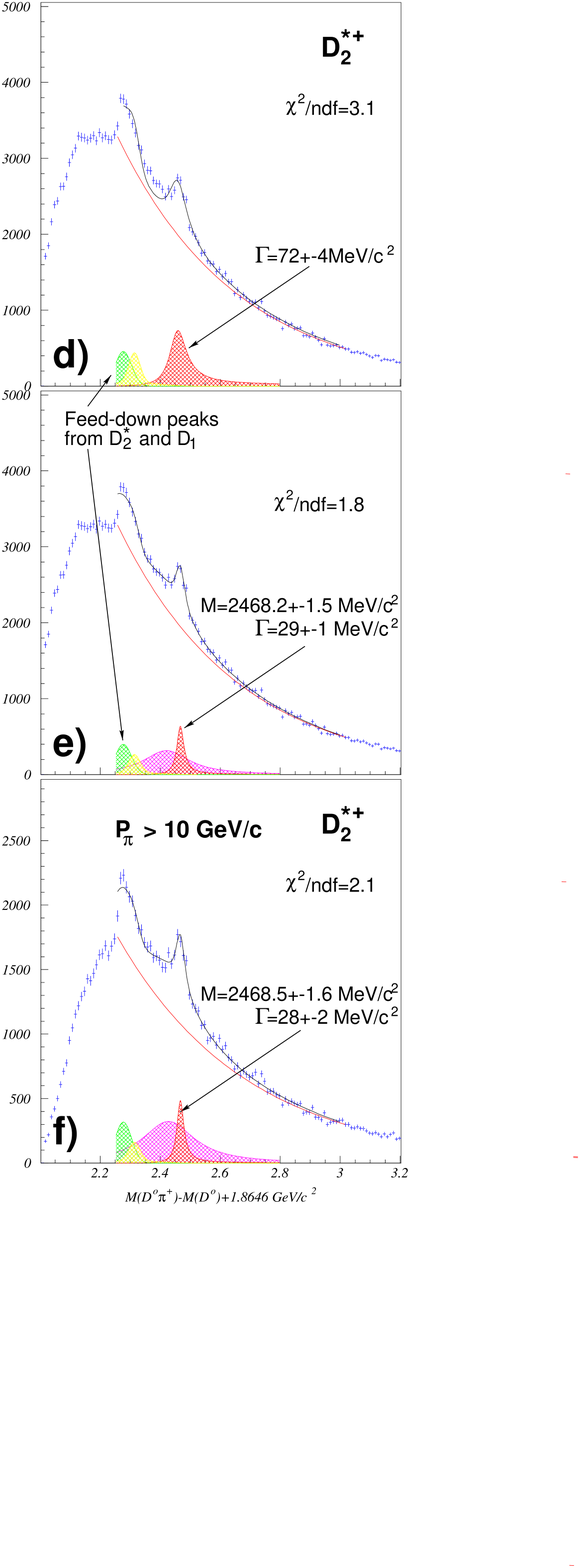}
\caption{
 The $D^+\pi^-$ ($D^0\pi^+$) mass spectra is shown in insets a, b, c (d, e, f).
 Insets c, f show the effect of a soft pion momentum cut.
\label{fig:d0+-}}
\end{figure}

The decays
 $D^0\rightarrow K^- \pi^+ $,
 $D^+\rightarrow K^- \pi^+\pi^+$,
 $D^0\rightarrow K^- \pi^+ \pi^- \pi^+$,
 $D^{*+} \rightarrow D^0 \pi^+$
 were selected\cite{fabbriosaka}. 
 The $D^+$ or $D^0$ candidates were combined with the pion tracks in the
 primary 
 vertex to form L=1 $D$-meson candidates in channels  $D^+\pi^-$ and
 $D^0\pi^+$. 
 Figure~\ref{fig:d0+-}a,b) shows the distribution in the invariant mass
 difference
 $\Delta M_0\equiv M(D^+\pi^-) - M(D^+) + M_{PDG}(D^+).$
 The plot shows a pronounced peak, consistent with being due to
 a $D_2^{*0}$ of mass  $M \approx 2460 \mev$. Because of the narrow width,
 this state has traditionally been identified as the $J=2^+$ state. The
 additional enhancement at $M\approx 2300 \mev$ is consistent, as verified
 from Monte Carlo simulations, with arising from the feed-down of the states
 $D_1^0$ and $D_2^{*0}$ decaying to $D^{*+}\pi^-$, with
 the $D^{*+}$ subsequently decaying to $D^+$ and undetected neutral pion.
 The $D_2^{*0}$ signal was fitted with a relativistic $D$-wave
 Breit-Wigner function, convoluted with a gaussian resolution function
 $(\sigma = 7\, {\rm MeV})$. The background was fitted with the  sum of an
 exponential, and two gaussians for the feed-downs described above, whose
 peaks and widths were fixed at the Monte Carlo values. The
 slope of the exponential was fixed to the value determined by a fit to the
 wrong-side events  mass distribution, which is very well described by a
 single-slope  exponential  in the entire fitting interval $2250-3000 \mev$.
 For this fit we get a $\chi^2/{\tt dof} =2$, and a $\Gamma = 55\pm 3 \mev$
 $D_2^{*0}$ width non compatible with the PDG2000 world
 average of $\Gamma = 23\pm 5 \mev$. We then add an  $S$-wave relativistic
 Breit-Wigner function  to the fit, which improves the fit quality
 $\chi^2/{\tt dof} =0.9$, and provides a width $\Gamma = 30\pm 2 \mev$
 compatible to the PDG2000 value.
 The mass difference
 $\Delta M_+\equiv M(D^0\pi^+) - M(D^0) + M_{PDG}(D^0)$
 spectrum (Fig.~\ref{fig:d0+-}d,e) shows structures similar to those in the
 $\Delta M_0$ spectrum, and  the fitting procedure follows the same
 guidelines. 
\par
 Several systematics checks have been
 performed to verify the stability of our measurements of masses and
 widths\cite{fabbriosaka}.
 As one example we show (Fig.\ref{fig:d0+-}c,f) how a $P>10\,{\rm GeV}/c$ 
 cut on the momentum of the soft pion provides peak and width
 statistically compatible.
 Table~\ref{tab:mass} summarizes the preliminary results on the
 measurements of masses and widths.
\par
 \begin{table}
 \tcaption{Preliminary measurements of masses and widths for narrow
 structures in   $D^+\pi^-$ and $D^0\pi^+$ invariant mass spectra. 
   \label{tab:mass}
 }
 \footnotesize
 \begin{center}
\begin{tabular}{l c c } 
             &  Mass                     & Width                 \\
             & $\mev$                    & $\mev$                \\
\hline
 $D_2^{*0}$  &  $2463.5 \pm 1.5 \pm 1.5$ & $30.5 \pm 1.9 \pm 3.8$ \\
  PDG2000    &  $2458.9 \pm 2.0$         & $23 \pm 5$             \\
\hline
 $D_2^{*+}$  &  $2468.2 \pm 1.5 \pm 1.4$ & $28.6 \pm 1.3 \pm 3.8$ \\
  PDG2000    &  $2459\pm4$               & $25^{+8}_{-7}$         \\
\hline
\end{tabular}
  \vfill
 \end{center}
\end{table}

\par
In conclusion, the study of the $D\pi$ mass
spectrum provides new preliminary values of the masses and widths for the
$D_2^*$ meson (Tab.~\ref{tab:mass}). The $D\pi$ mass spectrum (once
subtracted the background, the $D_2^*$ signal, and the expected feed-downs)
shows an excess of events centered around $2420 \mev$ and about $185 \mev$
wide. The observed excess could be reminiscent of the broad $D_0^*$
predicted by HQS, or of a feed-down from another broad state such as the $D_1
(j_q=1/2)$, possibly interfering.
Work is in progress to verify such hypothesis.  
\par
%
We wish to acknowledge the assistance of the staffs of Fermi National
Accelerator Laboratory, the INFN of Italy, and the physics departments
of the collaborating institutions. This research was supported in part
by the U.~S.  National Science Foundation, the U.~S. Department of
Energy, the Italian Istituto Nazionale di Fisica Nucleare and
Ministero dell'Universit\`a e della Ricerca Scientifica e Tecnologica,
the Brazilian Conselho Nacional de Desenvolvimento Cient\'{\i}fico e
Tecnol\'ogico, CONACyT-M\'exico, the Korean Ministry of Education, and
the Korean Science and Engineering Foundation. 
\par
\nonumsection{References}
\end{document}